\documentclass[a4paper,12pt]{article}
\pdfoutput=1
\usepackage[toc,page]{appendix}
\usepackage{mathtext}
\usepackage{misccorr}
\usepackage{indentfirst}
\usepackage{graphicx}
\usepackage{amsmath}
\usepackage{amsbsy}
\usepackage{textcomp}
\usepackage{amssymb}
\usepackage{upgreek}
\usepackage{rotating}
\usepackage{cite}
\usepackage{multirow}
\begin{document}
\title{Inner Shielding of the COMET Cosmic Veto System}
{\sf
\author{Oleg Markin\footnote{e-mail: markin@itep.ru}\\
{\small\emph{Institute for Theoretical and Experimental Physics,  Moscow 117218, Russia }}}
\date{\vspace{-5ex}}
\maketitle
\begin{abstract} A simulation of neutrons traversing a shield beneath the COMET scintillator strip cosmic-veto counter is accomplished using the Geant4 toolkit. A Geant4 application is written with an appropriate detector construction and a possible spectrum of neutron's energy. The response of scintillator strips to neutrons is studied in detail. A design  of the shield is optimized to ensure the time loss concerned  with fake veto signals  caused by neutrons from muon captures is tolerable.  Materials of shield layers are chosen, and optimum thicknesses of the layers are computed.   
\end{abstract}
}
\section{Introduction}
\begin{figure}
\centering
\includegraphics[width=.7 \textwidth]{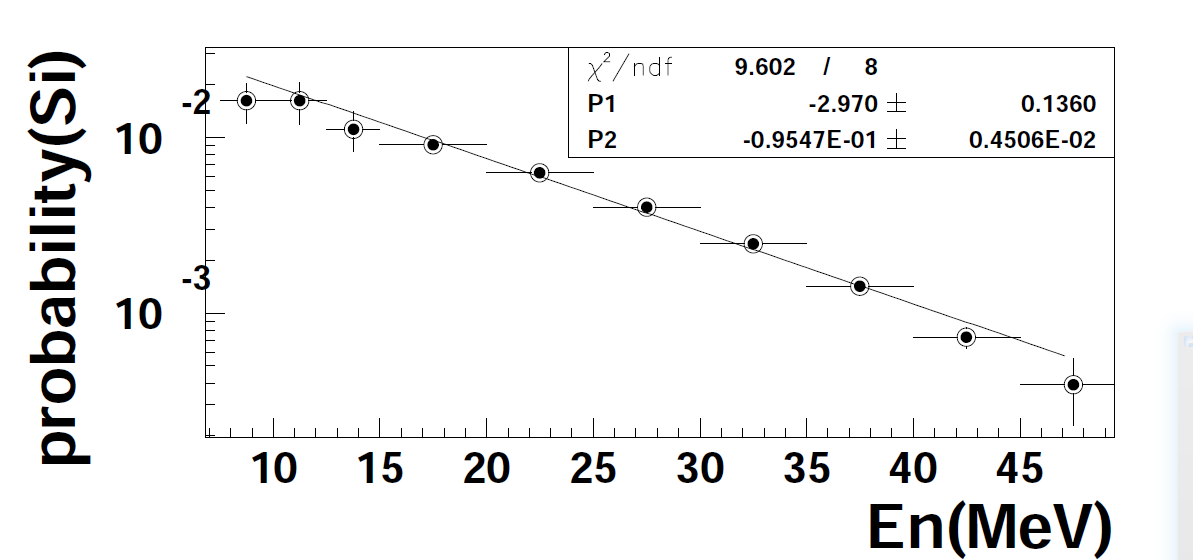}
\includegraphics[width=.7 \textwidth]{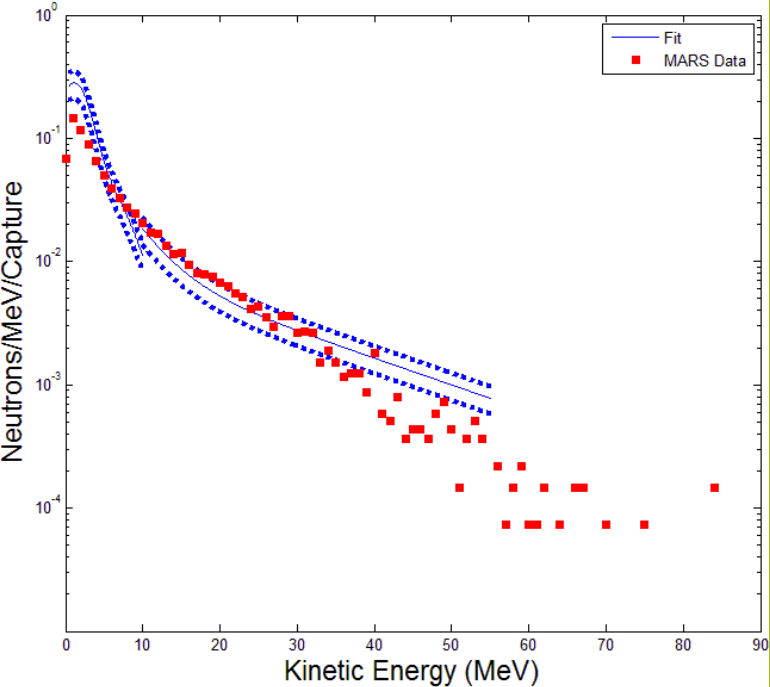}
\includegraphics[width=.7 \textwidth]{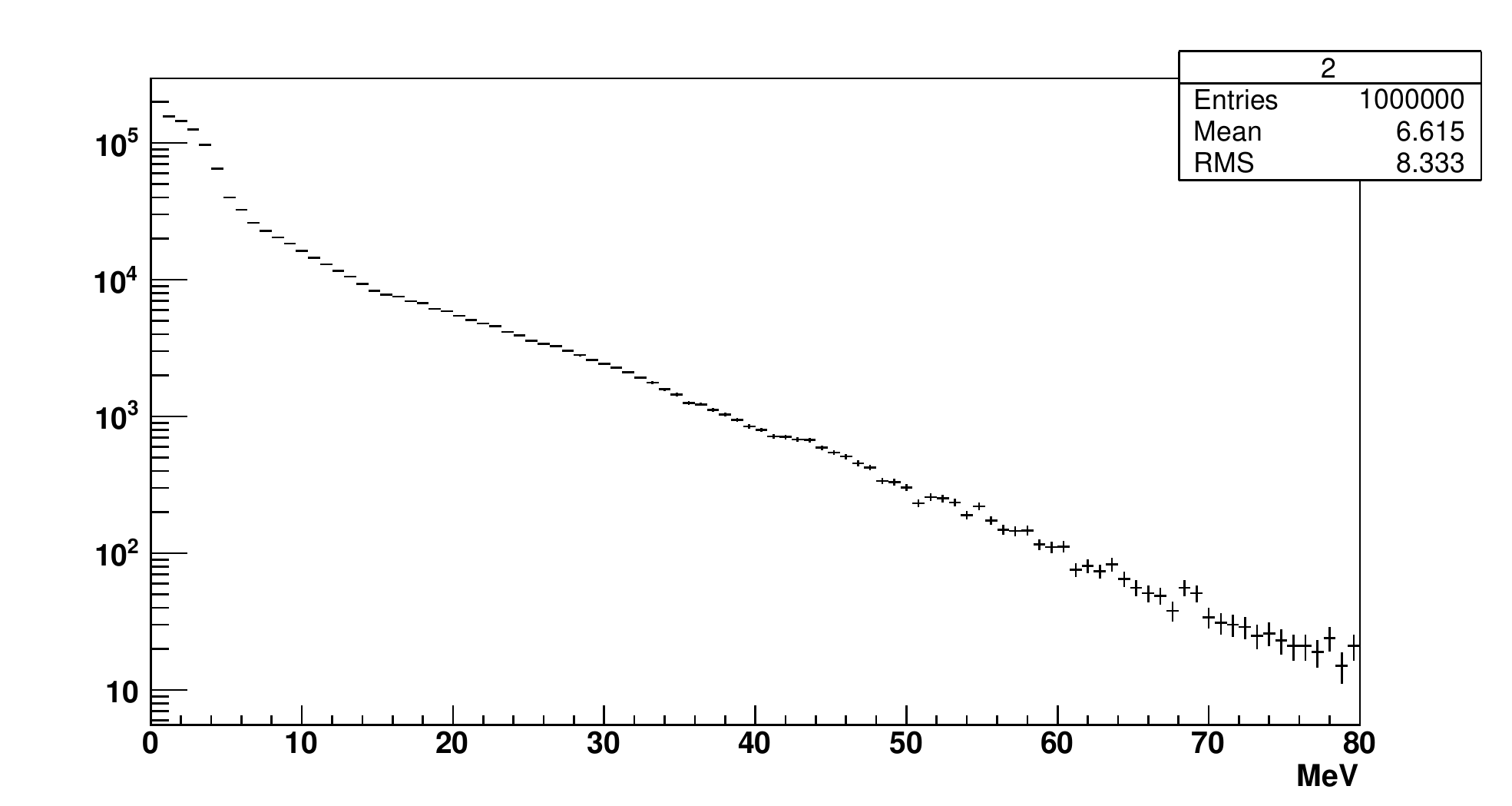}
\caption{ \textsf {Kinetic energy spectra of neutrons from muon captures in silicon \cite{Kozlowski} (top) and from an aluminum muon-stopping target \cite{Biliyar} (middle). Blue points and an associated fit come from an extrapolation of the measured calcium  spectrum to the aluminum one; red points are from a MARS calculation. Bottom: the simulated spectrum of neutrons from the muon-stopping target, which is used in this note and based on the spectra from works \cite{Kozlowski} and \cite{Biliyar}. }}
\label{nSpe}
\end{figure}

Neutrons  emitted  from the muon-stopping target  of the COMET experiment could cause fake veto signals in the Cosmic Veto Counter (CVC). Besides, they could damage silicon photo-multipliers MPPCs to be used in the CVC \cite{doc90}. On average, neutrons appear in six of ten  muon captures by nuclei of a target made of aluminum, that results  in  $\sim 10^9$ neutrons per second for the COMET Phase-I. Possible  spectra of neutrons from muon captures are shown in fig.~\ref{nSpe}. 

There are two major ways, in which neutrons affect the CVC: (i) through gammas from neutron captures by materials of the  COMET detector and (ii) kicking out protons and ions right in the CVC. The latter way is discussed  in detail in the section \ref{response}. Since there is a threshold for signal amplitudes of the CVC photo-detectors, signals of low-energy neutrons  are being rejected. Affordable values of the threshold lie from one tenth to one fifth of the average MIP signal in CVC, so the threshold corresponds to $0.2$--$0.4$ MeV of energy deposited in scintillator by ionization. Therefore,  we do not discuss low-energy neutrons in the section  \ref{response}, but only those with kinetic energy from 0.1 MeV to 1 MeV -- usually referred to as fast ones -- and medium-energy neutrons carrying up to 10 MeV. In the section \ref{shielding} we discuss energies typical for spectrum of muon captures. In all simulations presented in this note, the FTFP\_BERT\_HP physics list is used. 
 
\section{Strip response}
\label{response}
In order to clarify neutron's behavior in polystyrene, an ensemble of $10^6$ neutrons at normal incidence has been simulated. The energy threshold for the Geant4 proton creation has been set to 0.1 keV. The simulation has shown that fast and medium-energy neutrons traversing 7 mm thick polystyrene -- (C$_8$H$_8$)$_n$ --  scintillator interact predominantly elastically.
Those elastic interactions lead largely to production of charged projectiles -- protons and carbon ions -- that deposit their kinetic energy mostly through ionization causing scintillation in polystyrene. Track lengths of the charged projectiles lie in the sub-millimeter range. Their ionization energy loss per path length $dE/dx$ is much higher than that of MIP, so the resulting light yield per path length $dL/dx$ lessens due to saturation effects described by the Birks' empiric law \[dL/dx=L_0\frac{dE/dx}{1+k\cdot dE/dx} , \]  where $L_0$  is defined by the light yield at low ionization, and $k$ is a material-dependent coefficient that we take equal to 0.126 mm/MeV for polystyrene-based scintillators, following authors of \cite{Birks}. 

\begin{figure}[h]
\centering
\includegraphics[width=.7 \textwidth]{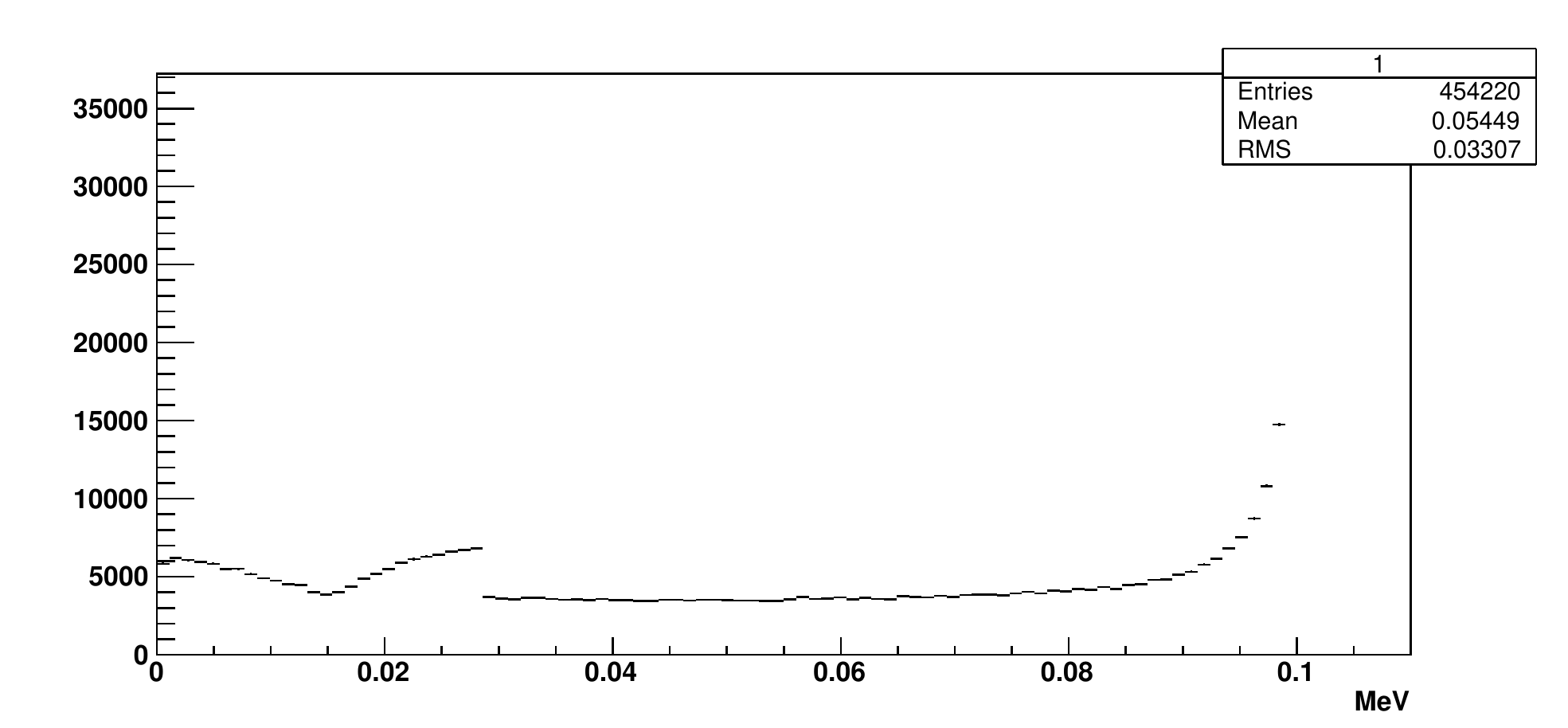}
\includegraphics[width=.7 \textwidth]{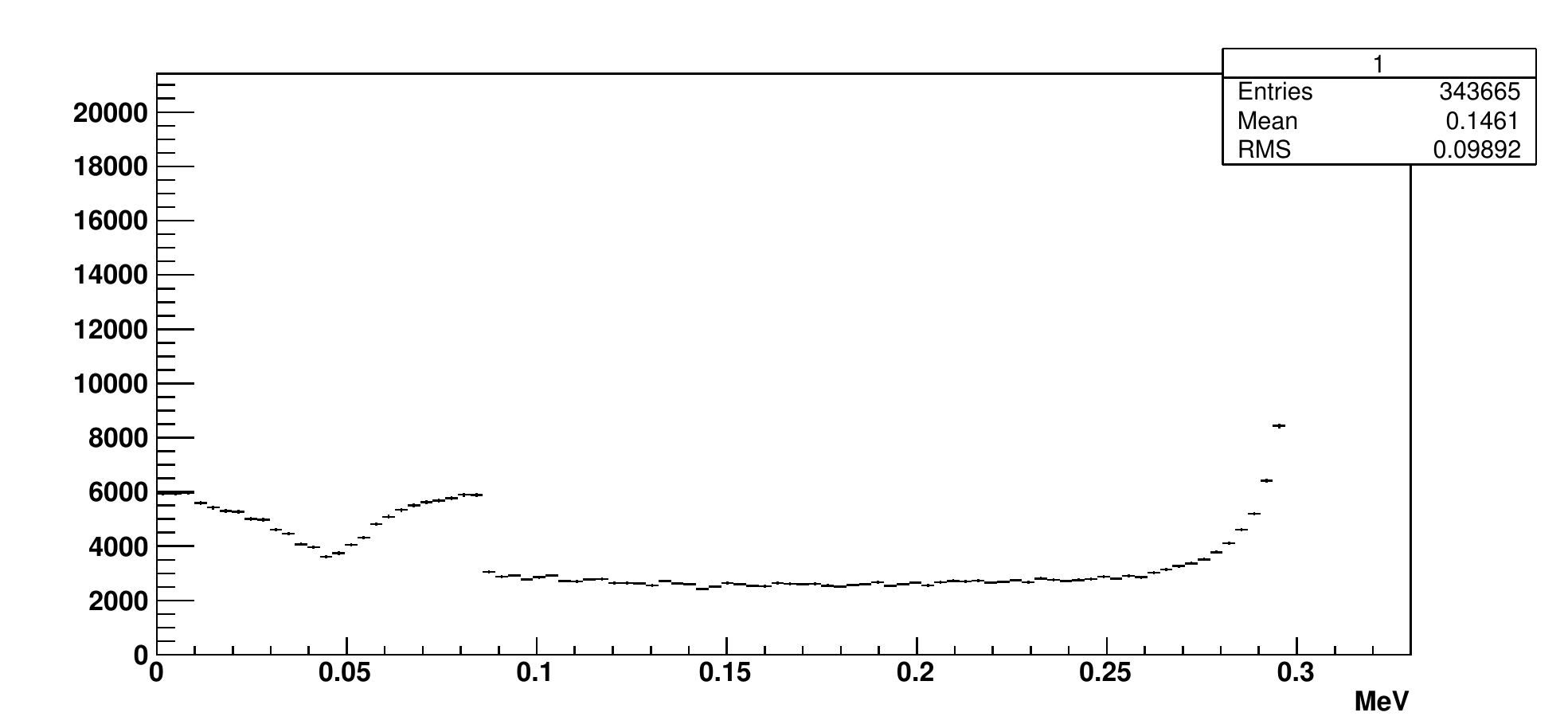}
\includegraphics[width=.7 \textwidth]{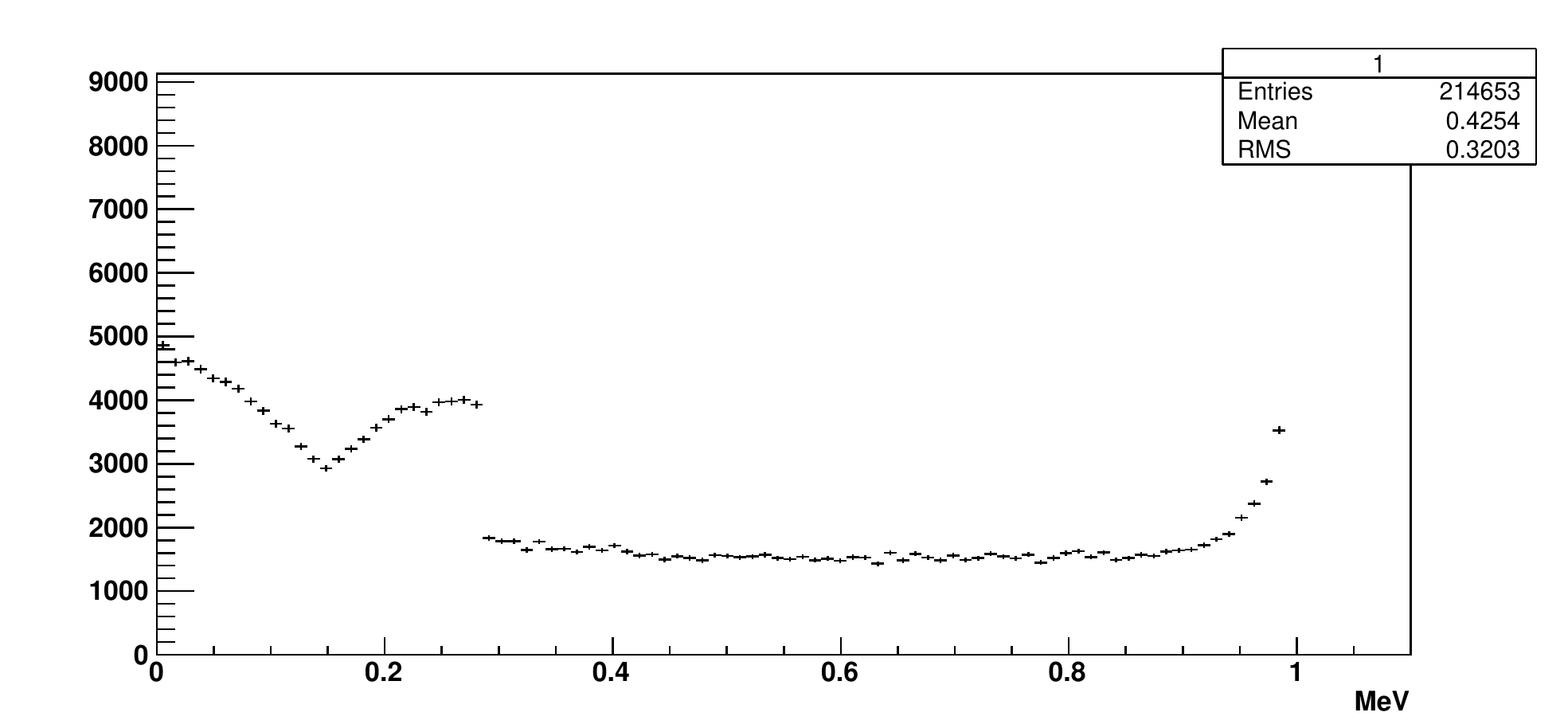}
\includegraphics[width=.7 \textwidth]{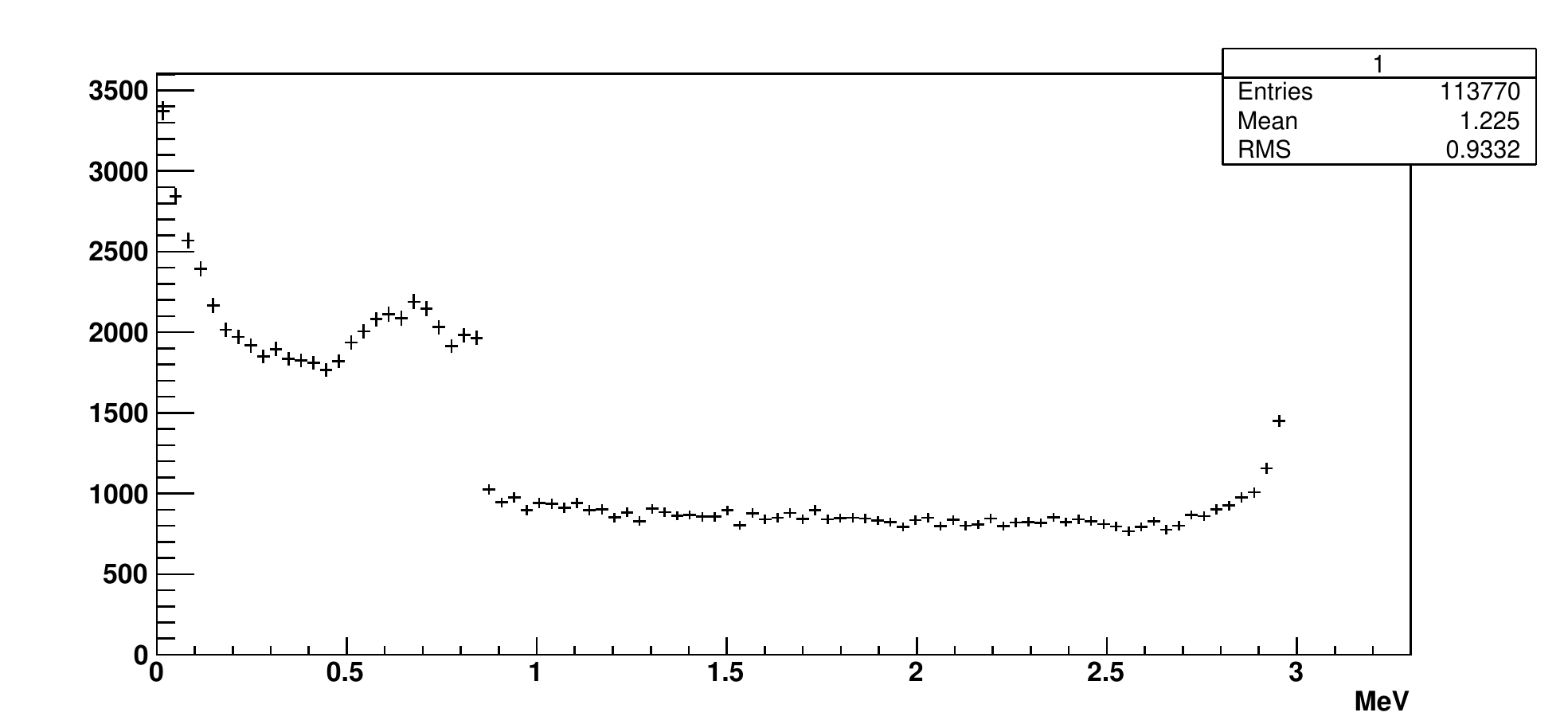}
\includegraphics[width=.7 \textwidth]{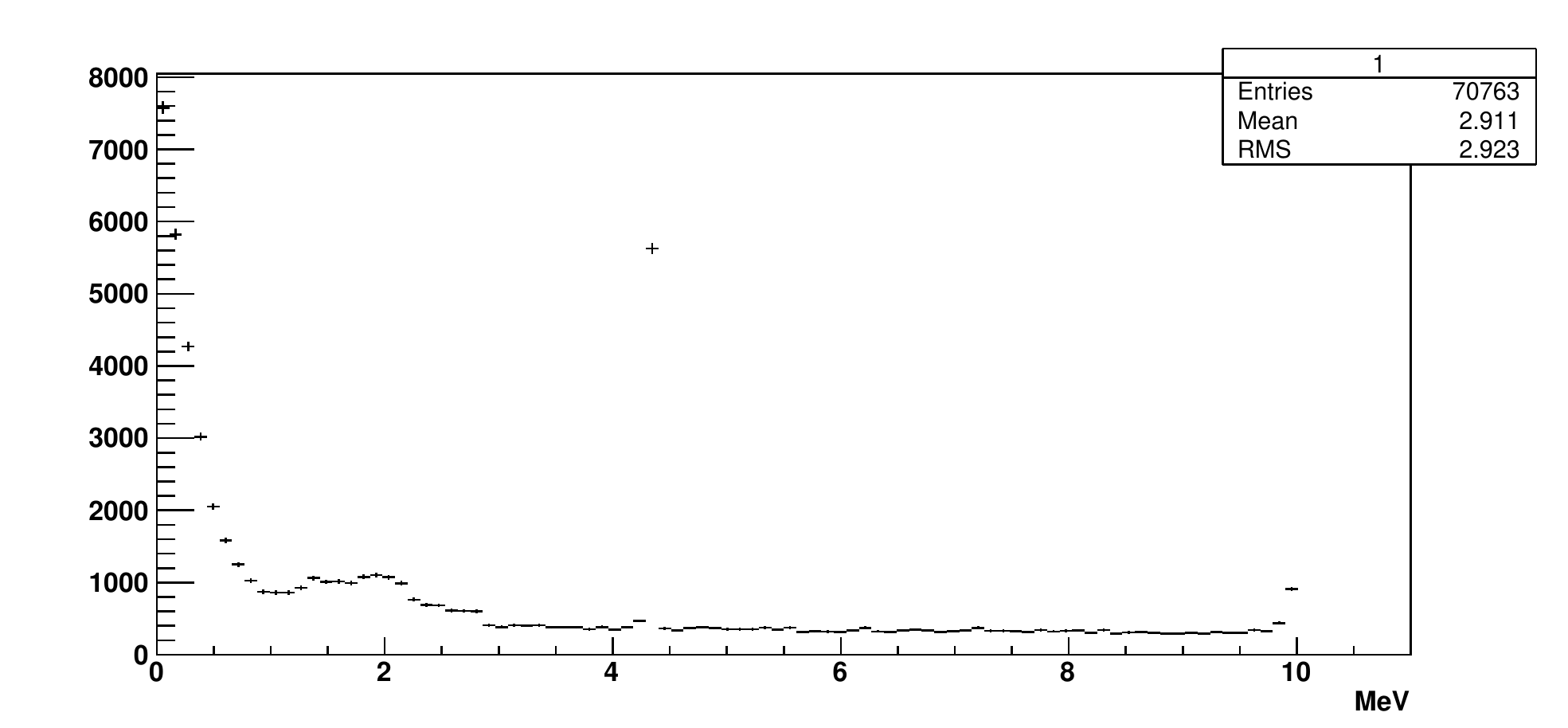}
\caption{ \textsf {Spectra of charged projectiles created by neutrons for energy of the neutrons equal to 0.1 MeV, 0.3 MeV, 1 MeV, 3  MeV and 10 MeV.}}
\label{speR}
\end{figure}

In order to quantitatively understand the Birks' effect for neutron's energies of interest, we have computed  spectra of the energy deposited to ionization by charged projectiles created in the elastic collisions  of incident neutrons, see fig.~\ref{speR}, as well as their track lengths. Table~\ref{estim} summarizes  mean values of both the energy loss and  the track length. The right column presents quotients of the two values, which are good estimates of the most probable value of the ionization energy loss per path length referred as stopping power.   
\begin{table}[ht]
\centering
\begin{tabular}[c]{|c|c|c|c|}\hline
  Energy  & Mean & Mean track length  &  Stopping \\
  of neutron, & ionization, & of charged projectiles,  & power,\\
   MeV &  MeV &  mm  & MeV/mm \\ \hline \hline 
  0.1  & 0.055 & 0.0011 & 50 \\ \hline 
  0.3 & 0.15 & 0.0019 & 79 \\ \hline
  1 & 0.43 & 0.0075 &  57 \\ \hline
  3 & 1.2 & 0.046 & 26 \\ \hline
  10 & 2.9 & 0.26 & 11 \\ \hline
\end{tabular}
\caption { \textsf {Estimation of stopping power for charged projectiles created by incident neutrons for five values of neutron's energy.}}
\label{estim}
\end{table}

The above estimation  shows that the stopping power for charged projectiles created by fast and medium-energy neutrons in polystyrene is much higher than that for MIP ($\simeq $ 0.19 MeV/mm). Using the Birks' law one can see that for neutron's energy up to 1 MeV, light yield of projectiles lies below the readout threshold values quoted in the introduction. Therefore, such neutrons are not typically seen by the CVC.
Furthermore, the sub-millimeter range of the charged projectiles in strips makes coincident signals in two  strips extremely unlikely.

\section{Shielding}
\label{shielding}
The most convenient shielding material by far is concrete, possibly doped with either boron or lithium to reduce gamma emission from neutron captures \cite{mu2e}. Concrete is comparatively chip, strong, available in bricks, and allows pouring, which  simplifies the construction of complicate profiles. It rather effectively shields against  low-energy and fast neutrons,  moderating  and then capturing them. However, the spectrum of neutrons from  muon captures stretches up to tens of MeV. It takes more elastic collisions to moderate energetic neutrons. In captures by light and intermediate weight nuclei,  cross sections for low-energy and fast neutrons is roughly proportional to the reciprocal of the neutron's speed  \cite{Rinard}. For these reasons, energetic neutrons have longer free path, i.e. higher penetration ability,  and this trend persists at higher energies. Furthermore, at higher energies, neutrons interact inelastically and cause spallation, giving birth to high-energy projectiles. Thus, for the COMET inner shielding a sophisticated  three-layer structure is necessary  \cite{CDR}.

\begin{figure}
\centering
\includegraphics[width=.6 \textwidth]{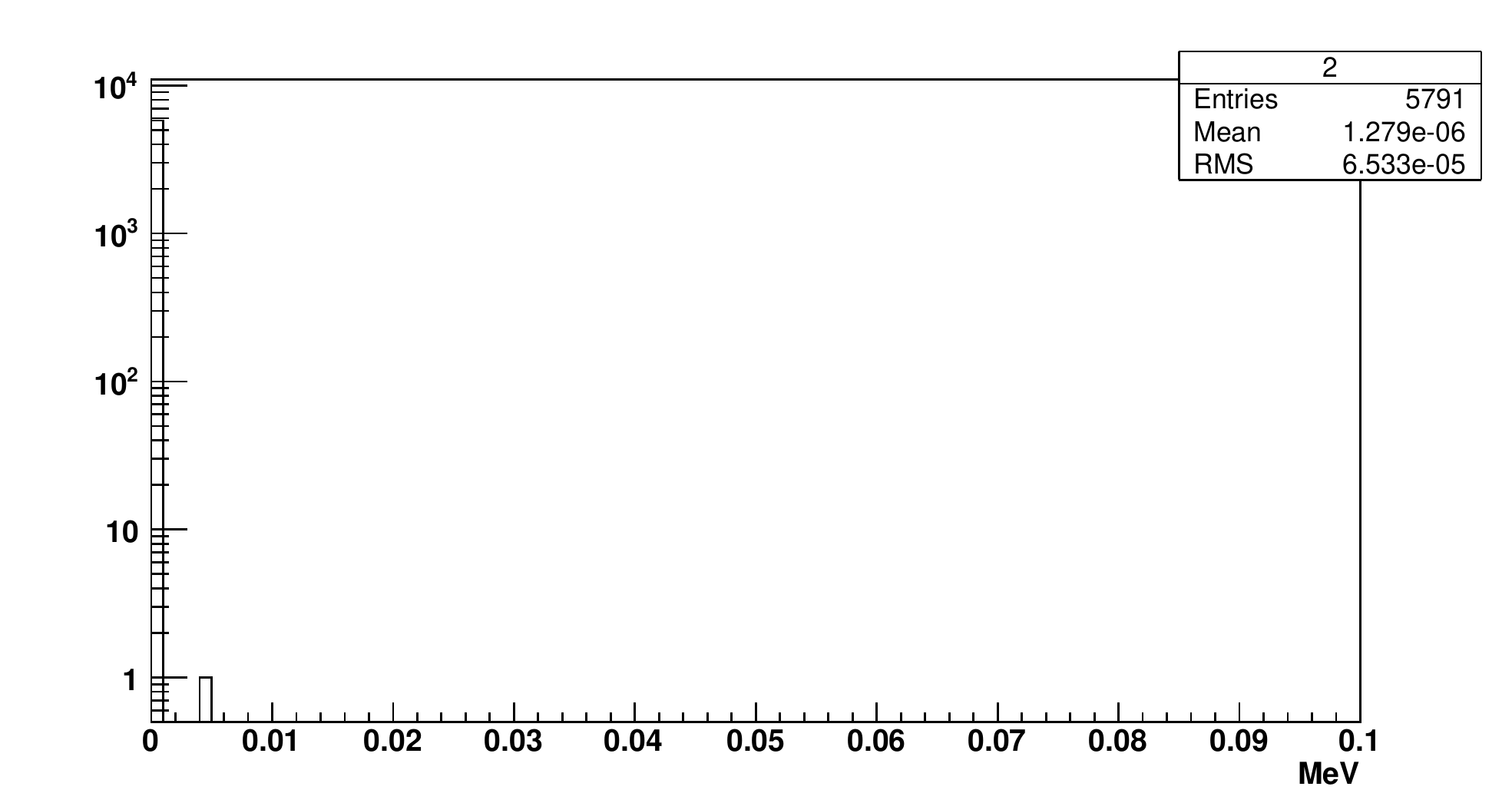}
\includegraphics[width=.6 \textwidth]{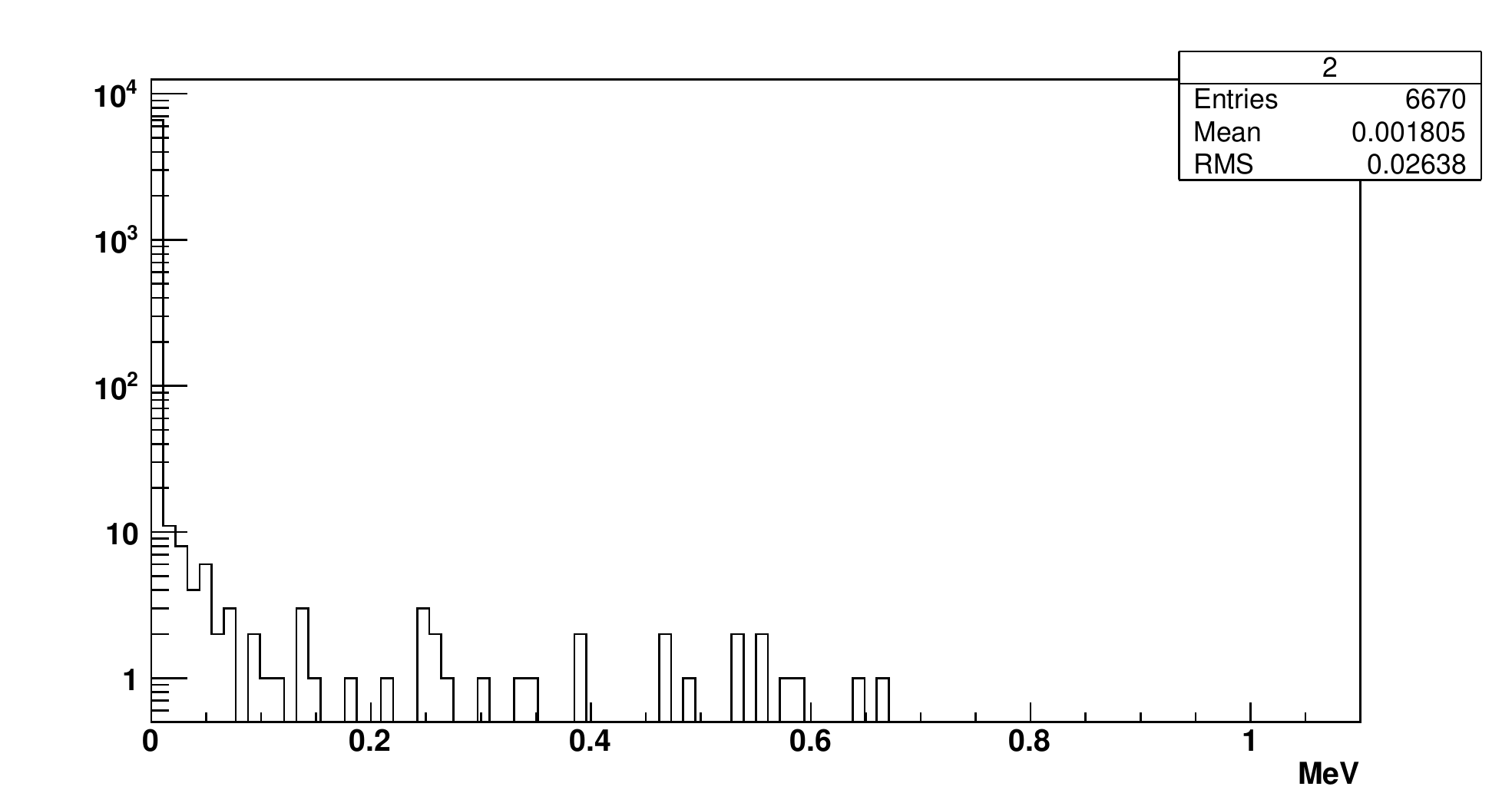}
\includegraphics[width=.6\textwidth]{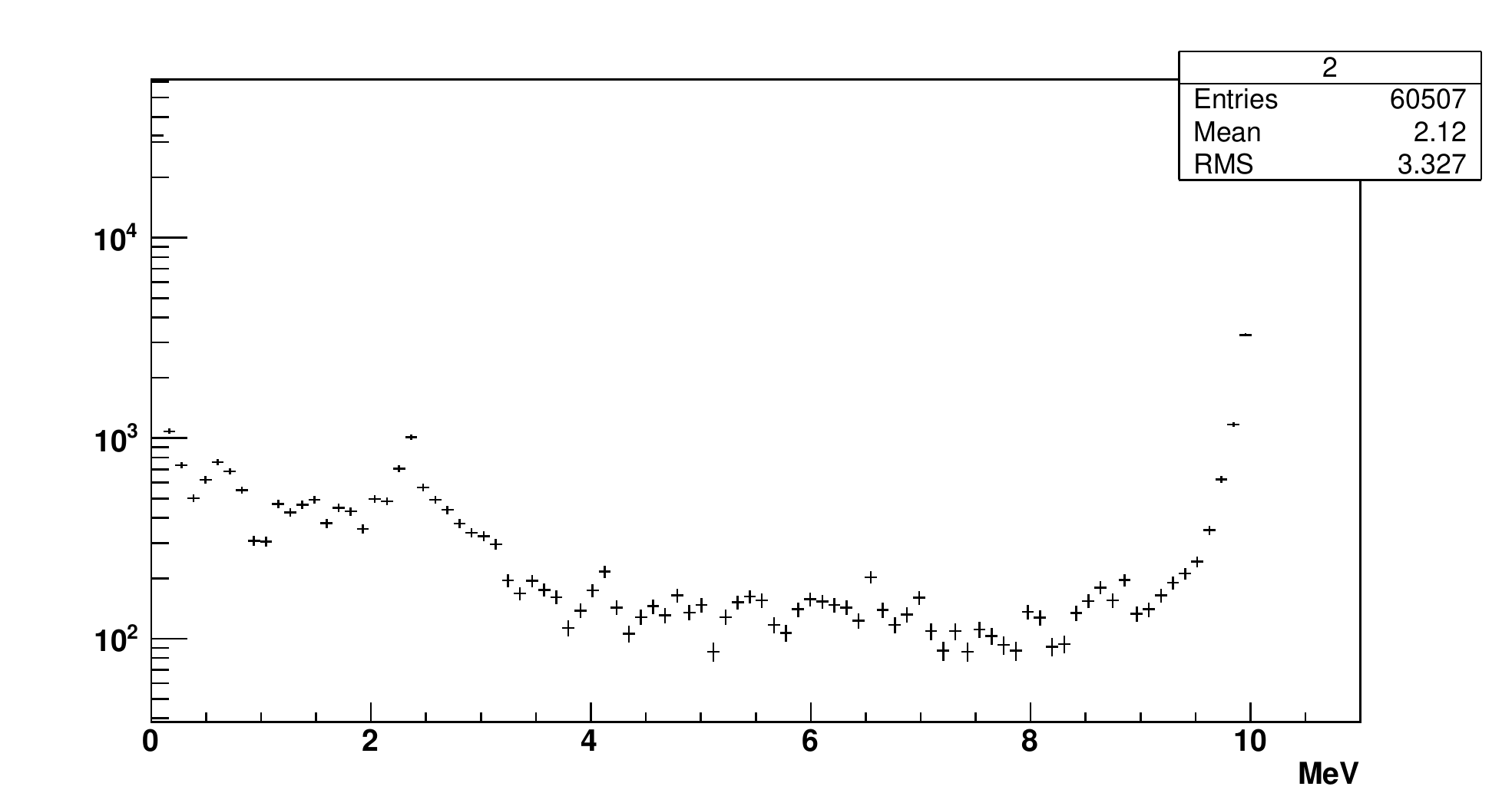}
\includegraphics[width=.6\textwidth]{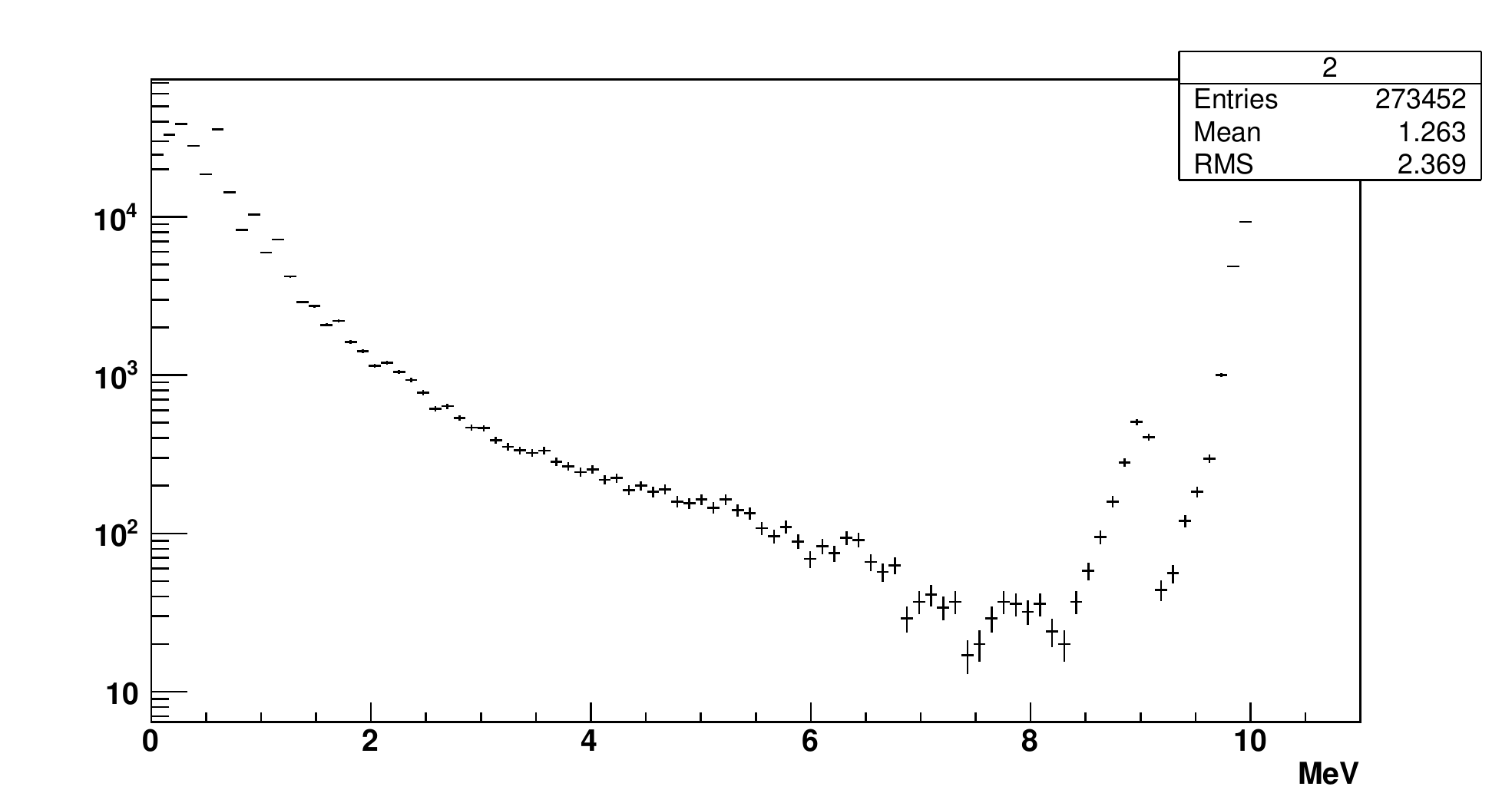}
\includegraphics[width=.6\textwidth]{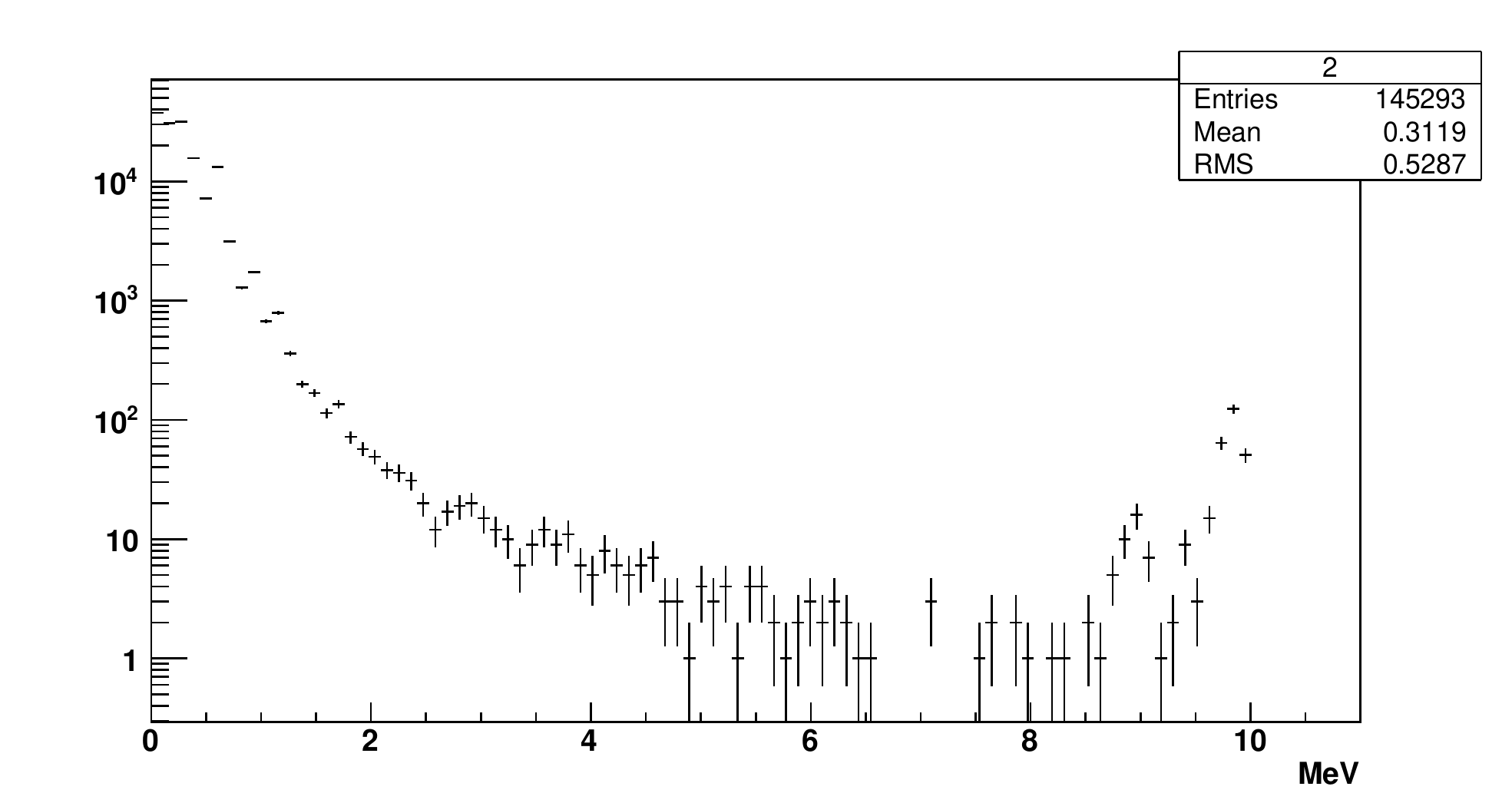}
\caption{ \textsf {Spectra of neutrons   (three top plots) behind the 50 cm thick concrete slab for energy of incident neutrons equal to 0.1 MeV, 1 MeV and 10 MeV, and (two bottom plots) behind 25 cm and 50 cm thick iron slabs for energy of incident neutrons equal to 10 MeV.}}
\label{concr}
\end{figure}

We have simulated 10$^6$ neutrons from a point source, impinging on 50 cm thick shielding slab of 4 $\times$ 4 m$^2$ area located 2 m far from the source. Figure~\ref{concr} shows neutron spectra behind a slab of concrete for different energies of incident neutrons. For 10 MeV neutrons, the neutron flux is only reduced by a modest factor 16 and includes a large fraction of medium-energy neutrons.  

Iron is not that good in construction of such a heavy self-supporting structure as the COMET inner shield.  Still, iron makes much better job in moderating  medium-energy neutrons, though at the cost of a sea of fast neutrons,  cf. bottom plots in fig.~\ref{concr}. Resulting fast neutrons can be  moderated quite effectively through elastic scattering in  polyethylene, which may however result in neutron captures on hydrogen accompanied by emission of 2.2~MeV gammas undesirable for the CVC. Gammas emitted in the neutron captures should be shielded by such a high-$Z$ material as lead. Thus, the three layers of different materials represent a better choice for the inner neutron shielding.   

In order to optimize the thickness of each layer, we have simulated 10$^6$ neutrons impinging on a three-layer slab of 4 $\times$ 4 m$^2$ area, with energy spectrum shown in the fig.~\ref{nSpe} bottom plot. Materials of layers are iron in the first one, polyethylene in the second one and lead in the third one.  Behind the slab,  there are a 10~mm thick air gap and then four 7~mm thick scintillator layers interleaved by 3~mm thick polystyrene pads.

We varied  thickness of the three shield layers, keeping the total shield thickness equal to 45~cm, and recorded spectra of neutrons and gammas in the air gap, as well as the response of each of the four scintillator layers. The energy deposited in scintillator by charged projectiles was corrected by the Birks' law. The corrected energy was transformed to the number of fired pixels of photo-detectors located at  left and right edges of each scintillator layer. Afterward, the number of the fired pixels was randomized with the Gauss distribution to account for fluctuations in the number of photons firing pixels.  Besides, we took into account the attenuation of light on its way to  photo-defector in such a way, as if the scintillator layers were subdivided  into 4~cm width strips, see  \cite{doc90}. At the end, we counted events where values of signals exceeded the threshold at least in two scintillator layers. 

\begin{figure}
\centering
\includegraphics[width=.67 \textwidth]{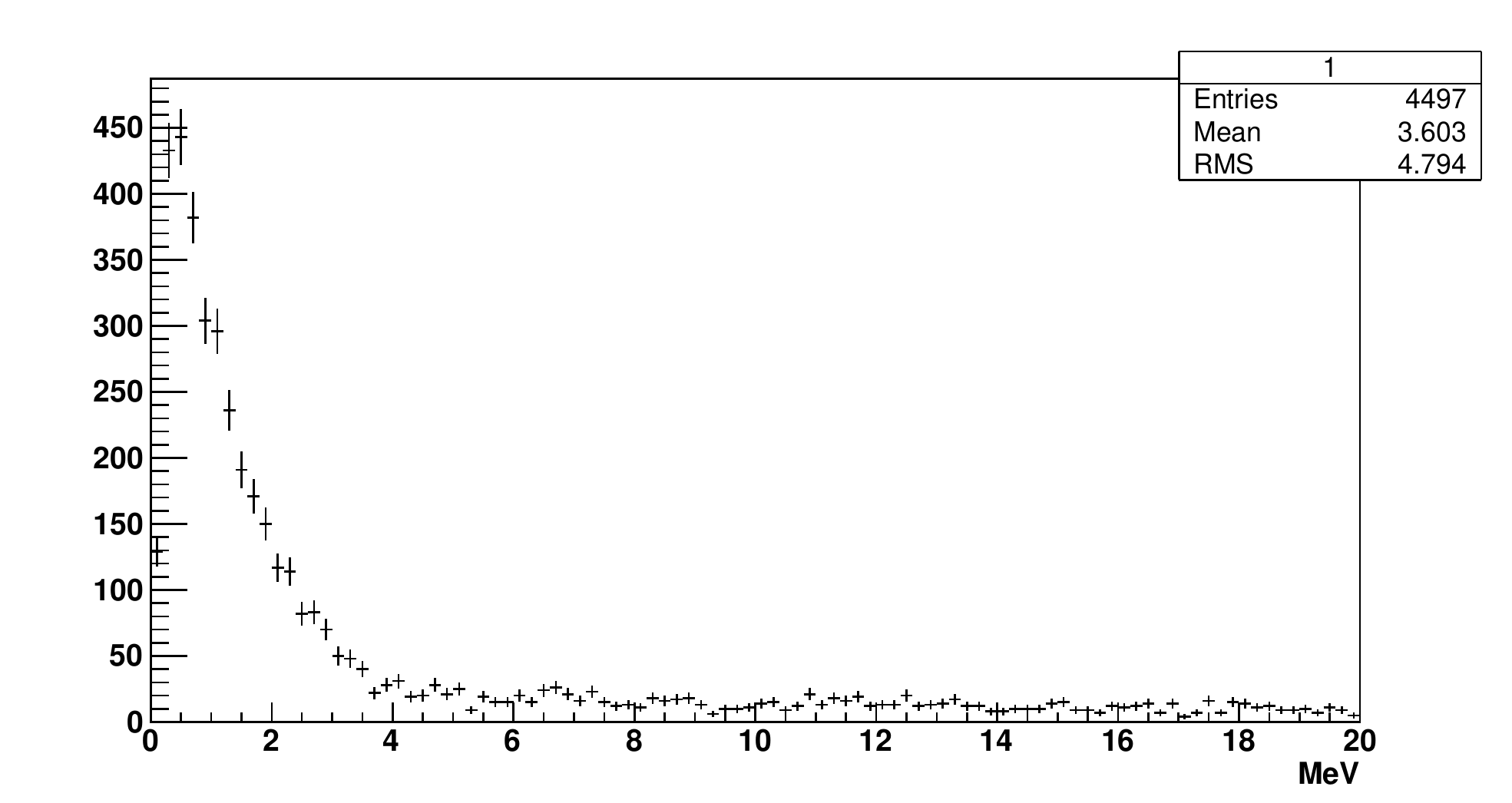}
\includegraphics[width=.67 \textwidth]{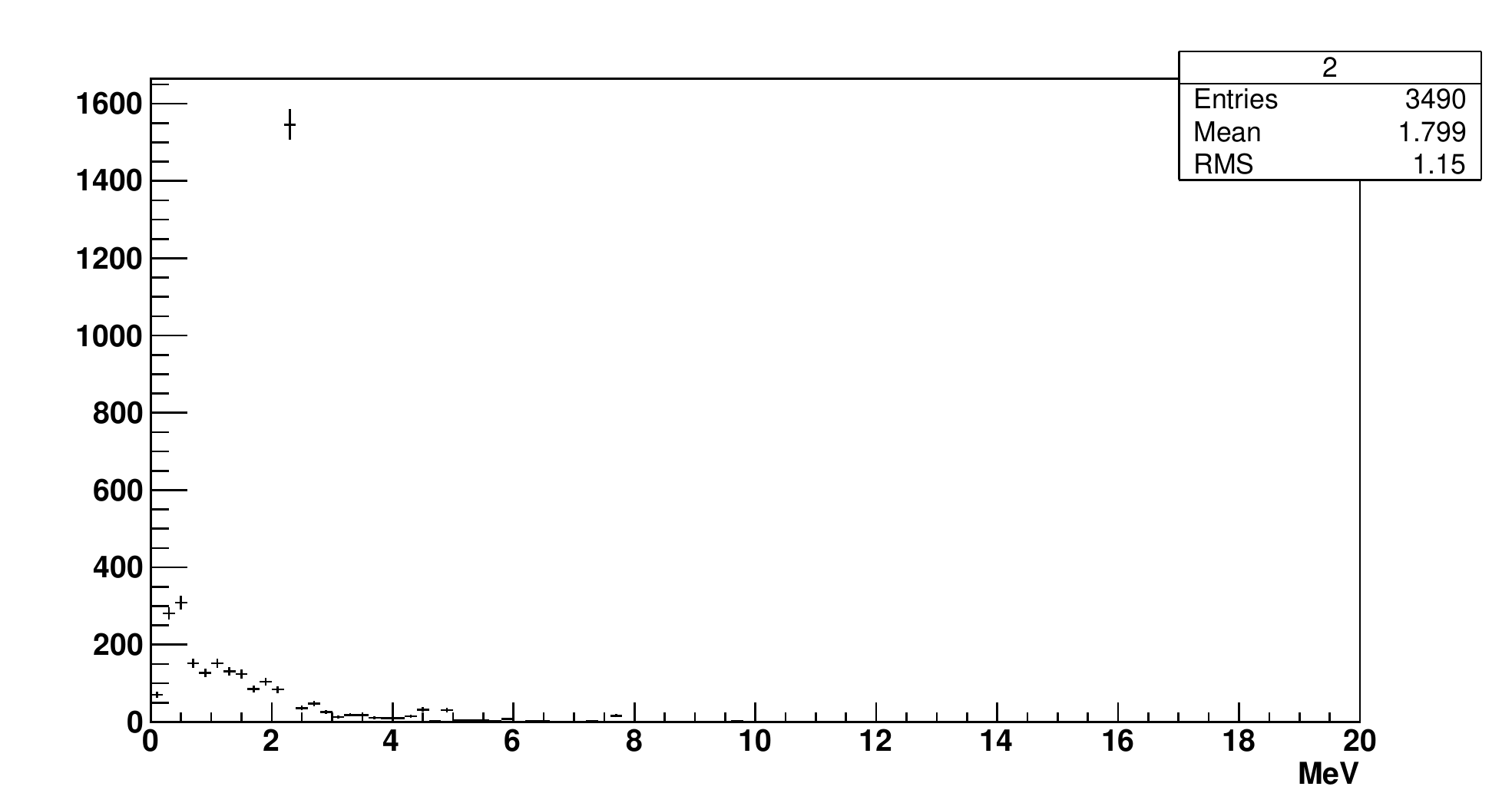}
\includegraphics[width=.67\textwidth]{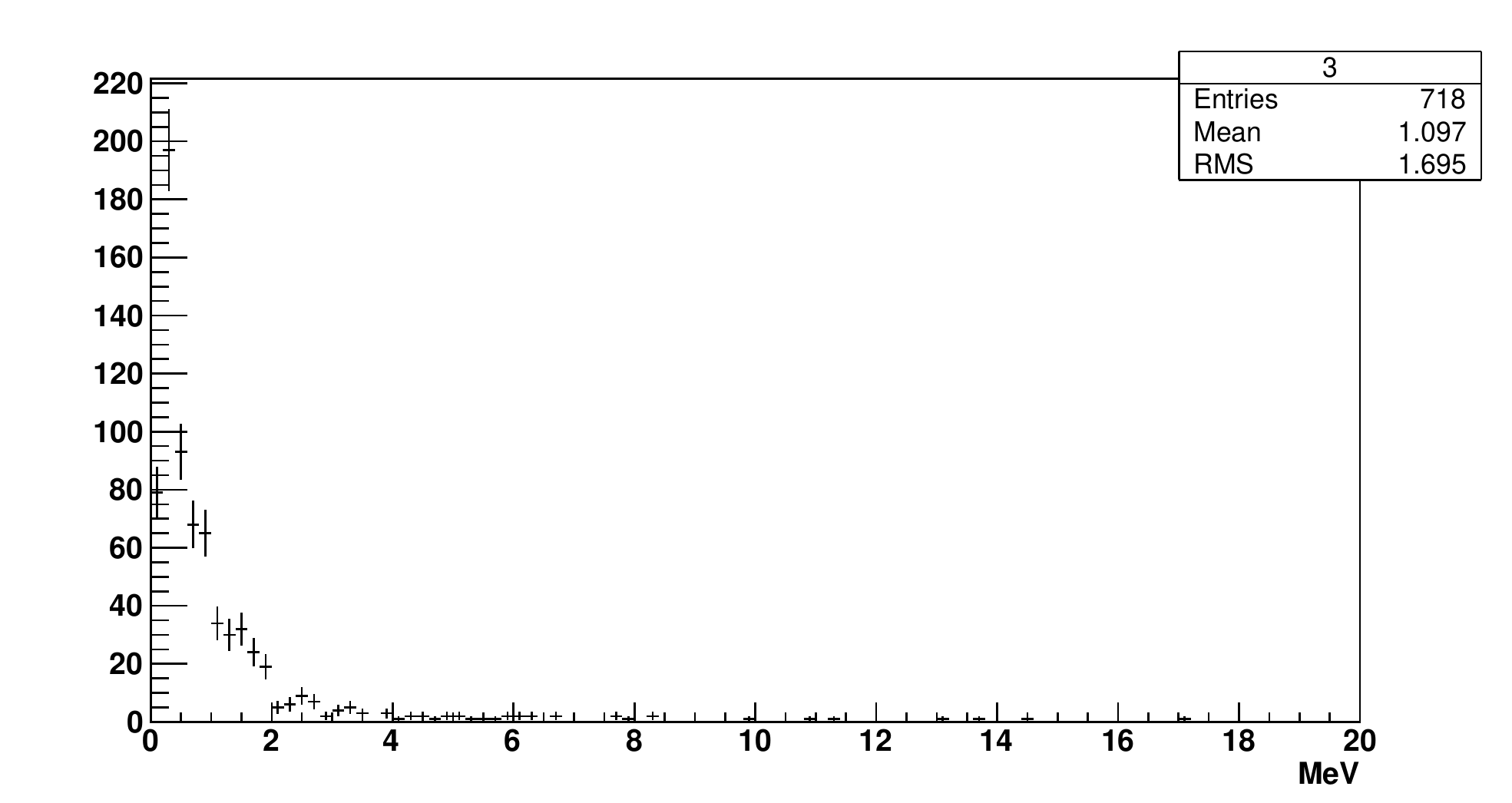}
\includegraphics[width=.67\textwidth]{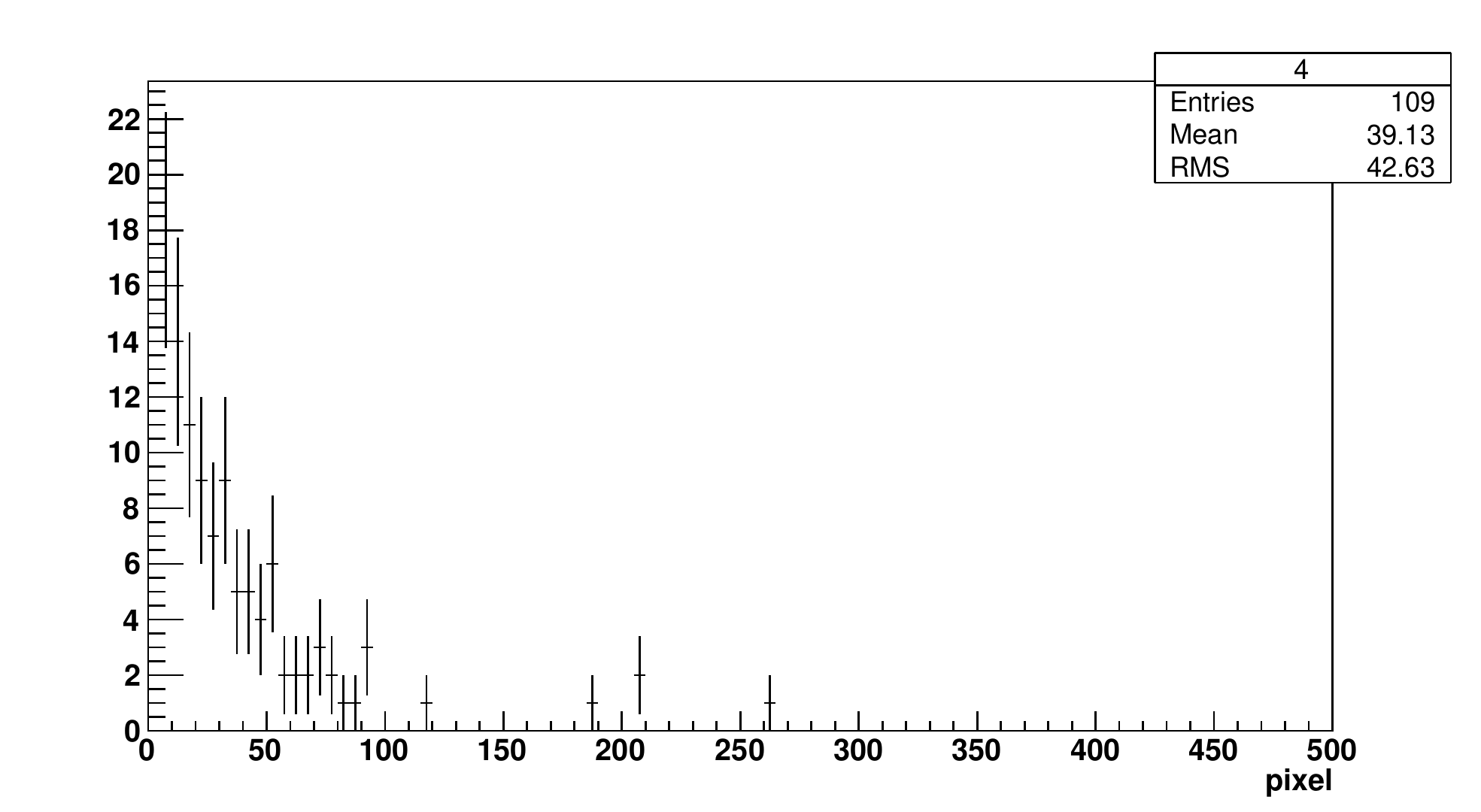}
\caption{ \textsf {Spectra of neutrons and  gammas in the 10~mm thick air gap behind the 45~cm thick (25~cm of iron + 10~cm of polyethylene + 10~cm of lead) shield (two top plots). Spectra of energy deposited to ionization, and  the number of fired pixels in the first of 4 scintillator layers behind the air gap for the case of a coincident signal in other layer(s) (two bottom plots). Events with energy below 0.2 MeV and signals below 5 pixels are omitted.}}
\label{coin}
\end{figure}

Table~\ref{thick} summarizes  the number  of events with coincident signals above either five or ten (written in brackets) photo-detector pixels. As an example, figure~\ref{coin} shows the spectra for combination of 25~cm of iron,  10~cm of polyethylene,  and 10~cm of lead. For this combination, the number of coincident signals above 5 pixels is equal to 240 that is $2.4\cdot 10^{-4}$ of the number of incident neutrons.  As the expected muon capture neutron rate at the shield slab is about $ 1.7\cdot 10^{8}$ per second, one gets about $ 4\cdot 10^{4}$ coincident signals per second in the scintillator layers behind the shield. Those signals fake cosmic-muon signals. Then, for the whole CVC with four side half-slabs, the rate of the fake signals is about 120 kHz. Thus, if we apply a 50 ns wide veto window, we have about 0.6\% of time lost due to neutrons from the muon captures. 

The use of aluminum instead of polystyrene as a pad material must be favorable for absorption of electrons produced by gammas. However, the simulation has shown that the use of aluminum pads does not almost change the overall shied performance with 5 cm of lead, and considerably deteriorates it for thicker lead layers, when more gammas are being absorbed in the shield at the cost of more neutrons behind it.  
\begin{table}[ht]
\centering
\begin{tabular}[c]{|c|c|c|c|c|}\hline 
  thickness of       & 5 cm         & 10 cm        &  15 cm       & 20 cm\\ 
   lead              &              &              &              & \\ \hline \hline 
   thickness of      &              &              &              & \\
   polyethylene      &              &              &              & \\ 
\multirow{2}*{5 cm}  &  257+211+98  &  185+155+81  &  168+133+58  & 199+144+67  \\ 
                     & (196+166+82) & (131+129+57) & (128+100+51) & (145+100+50)\\ \hline
\multirow{2}*{10 cm} &  170+143+86  &  109+76+55   &  168+108+36  & 165+137+70  \\ 
                     & (125+100+64) & (91+52+44)   & (123+78+27)  & (110+93+50) \\ \hline            
\multirow{2}*{15 cm} &  176+118+61  & 123+77+38    &  161+114+43  & \\    
                     & (131+88+46)  & (82+57+22)   & (119+80+30)  & \\ \hline
\multirow{2}*{20 cm} &  179+164+74  & 164+100+48   &              & \\    
                     & (124+123+60) & (104+75+32)  &              & \\ \hline                    
\end{tabular}
\caption { \textsf {For different combinations  of shield layer thicknesses, shown is the number of events, in which there are signals either in the first, or in the second, or in the third layer of scintillator, coincident with  a signal in other layer, provided  the number of fired pixels in each layer exceeds five (ten). The sum of the three numbers presents the total number of signals coinciding at least in two of four scintillator layers.}}
\label{thick}
\end{table}

The above estimation of the fake veto rate is conservative for three reasons. Firstly, signals of cosmic muons are higher than that of MIP that was used as a reference energy here. Therefore,  the value of the threshold in pixels corresponds to a  value of energy in MeV higher than we used in simulation. Secondly, among the events with coincident signals, there might be those with one signal being distant from another, which could not be caused by muons aimed to the detector and should therefore be disregarded. Thirdly, the moment of signals is not considered in the simulation, whereas two signals that do not come within a resolution window can not belong to a muon. The two latter factors have been accounted for in the next simulation, where signals in two layers have been considered caused by a muon only if they are  separated at most  by 20 ns time and by three strip widths across strip direction. Such a  value of time interval is just the worst time resolution possible in the experiment, whereas the three-strip transverse span allows the entire detector to be covered, cf.~\cite{doc90}. 

Iron shields against medium- and high-energy neutrons more effectively than concrete, though it is less convenient from the  point of view of engineering. A good compromise is the use of reinforced concrete containing scrap iron. Therefore, reinforced concrete as a material of the first shield layer has been implemented in the next simulation, volume fractions of iron and concrete being about 0.33 and 0.66 respectively. Additionally, the 10 cm thick return yoke of solenoid has been placed in front of the inner shield in the simulation. Table~\ref{feCon} shows the number of neutrons and hard photons behind the shield as  well as  the number of events, in which there are signals either in the first, or in the second, or in the third layer of scintillator, coincident with a signal in other layer, provided the number of fired pixels in each layer exceeds five. The sum of the three numbers presents the total number of signals co-occurring at least in two of four scintillator layers. All the numbers are shown for different thicknesses of the three shield layers.
\begin{table}[ht]
\centering
\begin{tabular}[c]{|c|c|c|c|c|c|}\hline
  Reinforced & Polyeth., & Lead, & \# of neutrons & \# of gammas    & \# of coincident \\
  concrete,  &               &       & behind shield & behind shield & signals in counters   \\
   cm        &  cm           &  cm   &               &               &                  \\ \hline
    25       &  10           &   5   & 9600          & 5800          & 123+69+50        \\ \hline 
    30       &   5           &   5   & 14000         & 5200          & 109+59+32        \\ \hline
    30       &  10           &   5   &  6000         & 4100          & 66+50+26         \\ \hline
    35       &  $-$          &   5   & 21400         & 4000          & 88+60+35         \\ \hline
    40       &  $-$          &   5   & 13000         & 2600          &  57+32+25        \\ \hline
    50       &  $-$          & $-$   & 6100          & 11300         &  174+104+103     \\ \hline
\end{tabular}
\caption { \textsf {Summary of neutron and gammas abundance behind a shield comprised of layers of reinforced concrete, polyethylene and lead, see text.}}
\label{feCon}
\end{table}

\begin{figure}[ht]
\centering
\includegraphics[width= 0.8\textwidth]{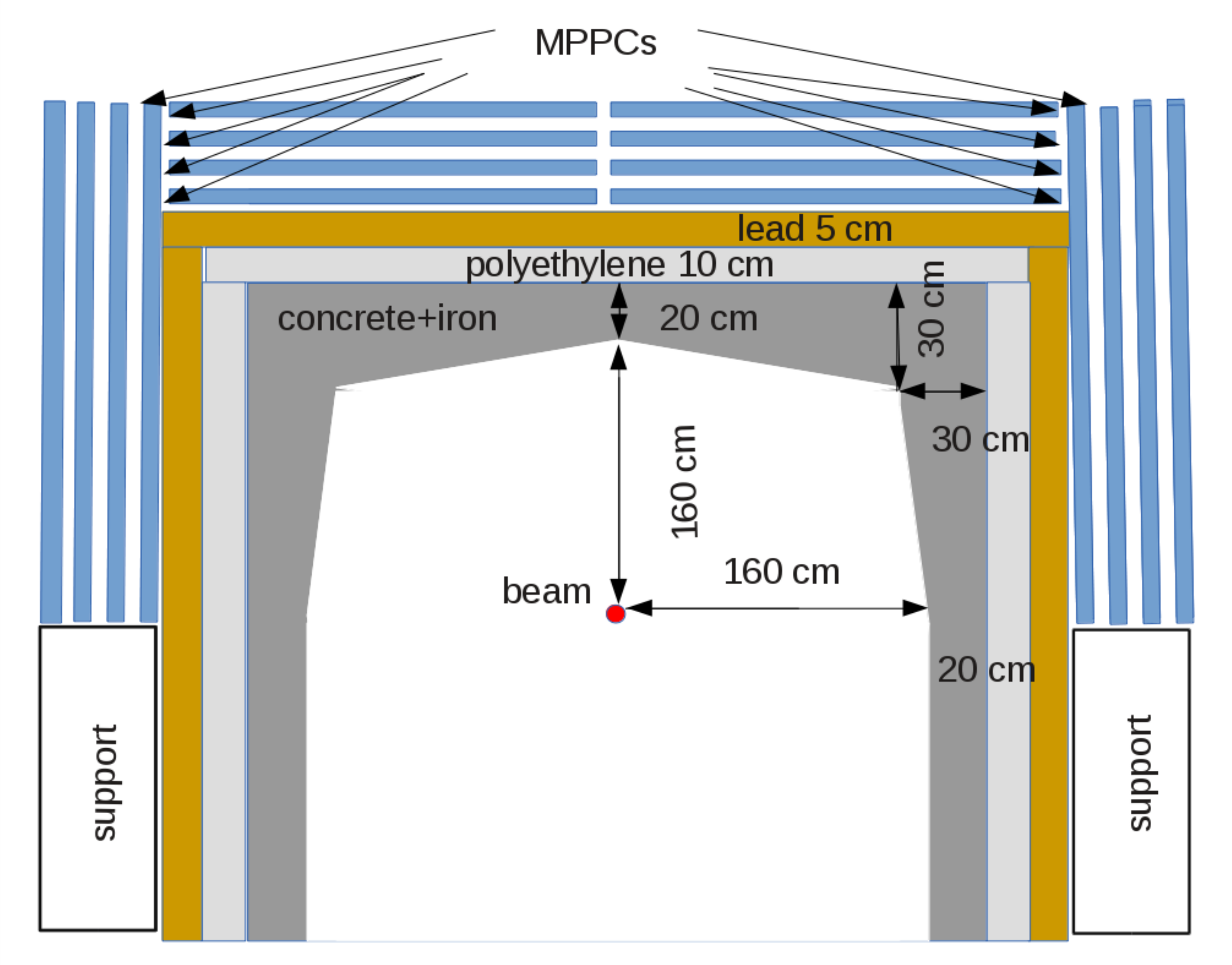}
\caption{ \textsf {A version of inner shield of arched shape (not to scale).}}
\label{shape}
\end{figure}

Dimensions of the shield depend on details of its composition and design. The optimal configuration of the shield is of arched shape, which accounts for signal attenuation along strips,  and better protects MPPCs that are located at outer ends of strips, see fig.~\ref{shape}. The final choice of shield's composition and shape shall be done taking into account the cost of materials, convenience of construction,  and affordable outer size the CVC.

\section{Conclusions}
Geant4 simulations have been performed including details of signal coincidence in a scintillator strip veto counter,  namely time and transverse spatial information from strips. Simulation have shown that a mix of iron and concrete  performs quite well as a base material for the inner shield of the CVC. In  the conjunction with the solenoid yoke, a shield composed of that mix, polyethylene and lead  drastically reduces the neutron flux at the counter strips. Thereby, the time loss concerned with fake veto signals of neutrons in the strips can be reduced to a tolerable value of 1\% of data-taking time by a 40--50 cm thick shield even at the 5-pixel threshold of photo-detectors. In the context of muon registration  efficiency, values of the photo-detector threshold up to 11 pixels can be safely used \cite{doc90}. That would additionally reduce the time lost due of fake veto signals by a factor up to 1.5, cf. table~\ref{thick}. The shield can be comprised of 20 to 30~cm of the iron-concrete mix, 10~cm of polyethylene and 5~cm of lead. Such a shield would reduce the flux  of fast and more energetic neutrons at photo-detectors  by 2 orders of magnitude. Since the expected value of the flux at CVC is about $10^3/$cm$^2$s, no substantial decrease of the photo-detector efficiency is expected throughout full period of data taking.    


\newpage
\begin{thebibliography}{99}
\bibitem{doc90}
O.~Markin and E.~Tarkovsky,
\emph{Simulations of the COMET veto counter}, COMET-doc-90, arXiv:1402.5522 [physics.ins-det]
\bibitem{Kozlowski}
T.~Kozlowski et al., 
\emph{Energy spectra and asymmetries of neutrons emitted after muon capture}, Nucl. Phys. A {\bf 436}, 717 (1985)
\bibitem{Biliyar}
V.~Biliyar,
\emph{ Calculation of the Spectrum of Ejected Neutrons from Muon
Capture}, Mu2e-doc-1619
\bibitem{Birks}
B.D.~Leverington, M.~Anelli, P.~Campana, R.~Rosellini, 
\emph{ A 1 mm Scintillating Fibre Tracker Readout by a Multi-anode Photomultiplier}, arXiv:1106.5649v2 [physics.ins-det]
\bibitem{mu2e}
The Mu2e Collaboration, 
\emph{Mu2e Conceptual Design Report }, Fermilab-TM-2545  (2012)
\bibitem{Rinard}
P.~Rinard,
\emph{ Neutron Interactions with  Matter}, in \emph{ Passive Nondestructive Assay of Nuclear Materials}, NUREG/CR-5550, LA-UR-90-732  (1991)
\bibitem{CDR}
The COMET Collaboration,
\emph{Conceptual Design Report for Experimental Search for Lepton Flavor Violating $\mu - e$ Conversion at Sensitivity of $10^{-16}$
with a Slow-Extracted Bunched Proton Beam (COMET)}, J-PARC P21 (2009)
\end{thebibliography}
\end{document}